\pdfoutput=1
\documentclass[aps,pre,twocolumn,superscriptaddress]{revtex4-2}

\usepackage{amsmath}
\usepackage{amssymb}
\usepackage{amsfonts}
\usepackage{graphicx}
\usepackage{hyperref}

\begin{document}


\title{Anti-persistent random walks in time-delayed systems}


\author{Tony Albers}
\email[]{tony.albers@physik.tu-chemnitz.de}
\author{David M\"uller-Bender}
\affiliation{Institute of Physics, Chemnitz University of Technology, 09107 Chemnitz, Germany}
\author{G\"unter Radons}
\affiliation{Institute of Physics, Chemnitz University of Technology, 09107 Chemnitz, Germany}
\affiliation{Institute of Mechatronics, 09126 Chemnitz, Germany}


\date{\today}

\begin{abstract}
We show that the occurrence of chaotic diffusion in a typical class of time-delayed systems with linear instantaneous and nonlinear delayed term can be well described by an anti-persistent random walk.
We numerically investigate the dependence of all relevant quantities characterizing the random walk on the strength of the nonlinearity and on the delay.
With the help of analytical considerations, we show that for a decreasing nonlinearity parameter the resulting dependence of the diffusion coefficient is well described by Markov processes of increasing order.
\end{abstract}


\maketitle

\section{\label{sec:I}Introduction}

Chaotic diffusion is a widely studied phenomenon in nonlinear dynamical systems, where the state variable shows diffusion.
It is well understood in low-dimensional systems such as low-dimensional Hamiltonian systems \cite{chirikov1979,lichtenberg1992,zacherl1986,geisel1987}
and one-dimensional iterated maps \cite{geisel1982,schell1982,fujisaka1982,geisel1985}, where the latter can be motivated by driven pendula, Josephson junctions, or phase-locked loops \cite{huberman1980,dhumieres1982}.
Beyond normal diffusion also anomalous diffusion, which is characterized by non-stationarity, nonergodicity, and infinite invariant measures, was extensively analyzed in such systems \cite{bel2006,metzler2014,albers2014,akimoto2015,albers2018,meyer2018}.
In contrast, there are only a few papers that consider chaotic diffusion in high-dimensional systems.
For instance, the works in \cite{cisternas2016,cisternas2018,albers2019_1,albers2019_2} consider chaotic diffusion of dissipative solitons in certain partial differential equation systems.
In this paper, we focus on another class of infinite-dimensional systems given by time-delay systems that are defined by delay differential equations (DDEs) \cite{hale1993,diekmann1995,hale2002},
which appear in all branches of science \cite{kuang1993,schoell2007,erneux2009,lakshmanan2011,schoell2016} and engineering \cite{erneux2009,stepan1989,michiels2007}.
While certain results can be inferred from the literature on diffusion in stochastic time-delay systems \cite{gushchin1999,budini2004,giuggioli_fokkerplanck_2016,loos2017,ando2017,geiss_brownian_2019},
there are only a few works on deterministic time-delay systems.
In \cite{wischert1994,schanz2003,sprott2007,dao2013_1,dao2013_2}, chaotic diffusion was observed in feedback loops with time-delay $\tau$ that are described by the DDE $\dot{x}(t)=\mu\sin\boldsymbol{(}x(t-\tau)\boldsymbol{)}$.
An integrated version of the Ikeda DDE \cite{ikeda1980} was considered in \cite{lei2011,mackey2021}.
Recently, we demonstrated that introducing a modulation of the time delay, i.e., $\tau=\tau(t)$, can lead to a giant increase of the diffusion constant over several orders of magnitudes \cite{albers2022},
which is associated with certain types of chaos induced by the time-varying delay \cite{mueller2018,mueller2019}.
In this paper, we show that, even if the delay is constant, chaotic diffusion in time-delay systems exhibits interesting features, where we focus on anti-persistence.
In general, anti-persistent random walks are characterized by negatively correlated increments, i.e., a step forward increases the probability that the next step is backwards and vice versa.
As a result, this leads to a reduction of the diffusion constant \cite{halpern_anti-persistent_1996}.
They can be observed, for instance, in the diffusion of charged particles \cite{maass_non-debye_1991}, in the dynamics of the basketball score during a game \cite{gabel_random_2012},
and in chaotic diffusion of dissipative solitons \cite{albers2019_1,albers2019_2}.
Anti-persistence is also present in fractional Brownian motion with Hurst exponent $H<1/2$ \cite{mandelbrot_fractional_1968},
which can be observed, for example, in crowded fluids \cite{ernst_fractional_2012}, albeit $H<1/2$ not necessarily implies anti-persistence in more general systems \cite{bassler_markov_2006}.
While it is known for stochastic systems that a time-delay can cause oscillations of the correlation function between positive and negative values \cite{ohira_oscillatory_1997}, to the knowledge of the authors,
anti-persistence in time-delay systems is not well understood, especially in the case of chaotic diffusion.

\section{\label{sec:II}Delay Equation}

We consider a typical class of delay differential equations (DDEs) with a linear instantaneous term and a nonlinear delayed term,
\begin{equation}
\label{eq:DDE}
\frac{1}{\Theta}\dot{x}(t)=-x(t)+f\boldsymbol{(}x(t-1)\boldsymbol{)},
\end{equation}
where the parameter $\Theta$ sets the time scale, and $f(x)$ is a nonlinear function.
Different choices of the nonlinearity $f$ lead to several time-delayed systems well known in the literature.
For instance, for $f(x)=\mu x/(1+x^{10})$, one obtains the Mackey-Glass equation \cite{mackey1977} describing the time evolution of the concentration of white blood cells,
whereas for $f(x)=\mu\sin(x)$, one gets the Ikeda equation \cite{ikeda1980,ikeda1982} describing the dynamics of the transmitted light from an optical ring cavity system,
where the nonlinearity is similar to the one in models for certain opto-electronic oscillators \cite{larger2013,chembo2019}.
There are several other nonlinearities that have been investigated \cite{lakshmanan2011}.
The time scale transformation $t'=\Theta t$ transforms Eq.~(\ref{eq:DDE}) to the DDE $\dot{y}(t')=-y(t')+f\boldsymbol{(}y(t'-\Theta)\boldsymbol{)}$ with $y(t')=x(t'/\Theta)$
demonstrating that large values of $\Theta$ correspond to the large delay limit.
Since we consider $\Theta\gg1$ in this work, our results contribute to the highly developed theory of singularly perturbed DDEs and systems with large delay
(cf. \cite{ikeda1982,chow1983,mallet-paret1986,ikeda1987,mensour1998,adhikari2008,amil2015,wolfrum2006,wolfrum2010,lichtner2011,giacomelli2012,marino2014,faggian2018,marino2019}).
In this article, we investigate nonlinearities $f$ for which the corresponding iterated map $z_{t+1}=f(z_t)$ is known to show chaotic diffusion \cite{geisel1982,schell1982,fujisaka1982,geisel1985}.
More specifically, we conser maps with reflection $f(-x)=-f(x)$ and discrete translational symmetry $f(x+1)=f(x)+1$.
It was shown that for sufficient damping, differential equations describing Josephson junctions, phase-locked loops, or driven damped pendula \cite{huberman1980,dhumieres1982} can be reduced to such iterated maps \cite{geisel1982}.
A paradigmatic example is the climbing-sine map given by
\begin{equation}
\label{eq:climbing-sine}
f(x)=x+\mu\sin(2\pi x),
\end{equation}
which shows chaotic diffusion for $\mu>\mu_c=0.732644...$ \cite{geisel1982}.
In a previous article \cite{albers2022}, we discussed that the resulting DDE, Eq.~(\ref{eq:DDE}) with Eq.~(\ref{eq:climbing-sine}), leads to chaotic diffusion for large enough $\Theta$,
where the state variable $x$ can thereby be interpreted as an unbounded phase variable.
In this article, we show that the diffusion process is well described by an anti-persistent random walk.
Although the following numerical results were all obtained for the climbing-sine nonlinearity, our qualitative findings, however, are general in so far as we checked that they occur also for other nonlinearities
such as the iterated map studied by Klages et. al. \cite{klages1995} or the climbing tent map \cite{fujisaka1982} in a wide range of parameters.

\begin{figure}
\includegraphics[width=\linewidth]{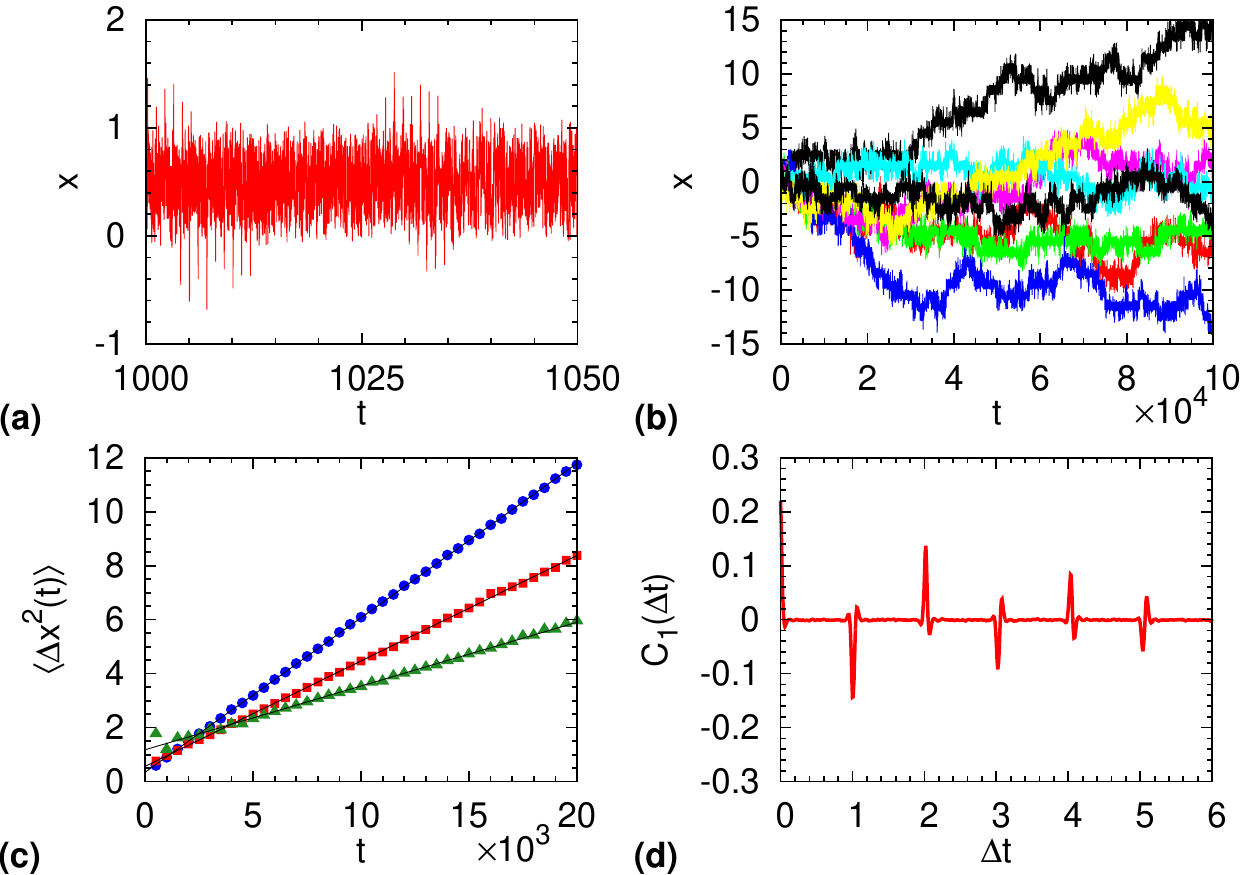}
\caption{\label{fig:trajectories}(a) A single solution $x(t)$ of the DDE, Eq.~(\ref{eq:DDE}), on a short time scale shows strong oscillations typical for turbulent chaos.
(b) On a larger time scale, an ensemble of solutions spreads diffusively reminiscent of Brownian motion.
(c) The mean-squared displacements for different values of $\Theta$ numerically obtained from $N=10^4$ trajectories of duration $T=2\cdot 10^{4}$ increase linearly ($\Theta=25,50,100$ from top to bottom).
(d) The covariance function $C_1(\Delta t)=\langle\delta x_1(t)\delta x_1(t+\Delta t)\rangle$ of the increments $\delta x_1(t)=x(t+1)-x(t)$ shows peaks of alternating algebraic sign, clearly demonstrating antipersistence.
Parameters of the simulations are $\Theta=50$ and $\mu=0.9$.}
\end{figure}

Equation~(\ref{eq:DDE}) can be formally solved by the method of steps \cite{bellman1965} leading to an iteration of solution segments $x_n(t)$ defined on time intervals $[n-1,n]$
given an initial function $x_0(t)$ on the time interval $[-1,0]$ \cite{ikeda1987},
\begin{equation}
\label{eq:solution}
x_{n+1}(t)=x_n(n)e^{-\Theta(t-n)}+\int_n^t\Theta e^{-\Theta(t-t')}f\boldsymbol{(}x_n(t'-1)\boldsymbol{)}\,dt'.
\end{equation}
This equation shows that for large values of $\Theta$, states $x$ for instants of time on the interval $[n-1,n]$ are mapped to the subsequent time interval by the action of the nonlinearity $f$
and then are smoothed by the kernel $\Theta\exp(-\Theta(t-t'))$ of width $1/\Theta$.
We numerically solved Eq.~(\ref{eq:DDE}) using the two-stage Lobatto IIIC method with linear interpolation \cite{bellen2003} and a step width $\Delta t=0.001$.
A typical solution of Eq.~(\ref{eq:DDE}) on a short time scale is depicted in Fig.~\ref{fig:trajectories}(a) and shows strong fluctuations of width $1/\Theta$ due to the chaos generating map $f$ and the smoothing kernel.
If we consider an ensemble of solutions of Eq.~(\ref{eq:DDE}) on a large time scale shown in Fig.~\ref{fig:trajectories}(b), we observe a diffusive spread of the trajectories that is reminiscent of Brownian motion.
In order to check whether this spread follows the laws of normal diffusion, we calculate the mean-squared displacement (MSD) $\langle\Delta x^2(t)\rangle$ defined by $\langle\Delta x^2(t)\rangle=\langle[x(t)-x(0)]^2\rangle$,
where the angle brackets denote an ensemble average over many solutions of Eq.~(\ref{eq:DDE}) with slightly different initial functions.
Fig.~\ref{fig:trajectories}(c) shows the numerically determined MSDs for different values of the parameter $\Theta$.
They all have in common a linear increase in time typical for normal diffusion, where the slope of the MSD defines the diffusion coefficient $D\simeq\langle\Delta x^2(t)\rangle/t$.
In order to understand the origin of the diffusion process from a microscopic point of view, one typically considers statistics of the increments of the process.
Here, we define an increment by $\delta x_{\eta}(t)=x(t+\eta)-x(t)$.
A first natural choice is $\eta=1$ due to the method of steps.
The covariance function $C_{\eta}(\Delta t)$ of the increments is defined by $C_{\eta}(\Delta t)=\langle\delta x_{\eta}(t)\delta x_{\eta}(t+\Delta t)\rangle$.
Here, we assumed that the covariance function is stationary, i.e., does not depend on $t$, what can be expected from the time-translational invariance of Eq.~(\ref{eq:DDE}) and was in addition confirmed numerically.
The numerically determined covariance function of the increments for $\eta=1$ is shown in Fig.~\ref{fig:trajectories}(d).
It consists of peaks of alternating algebraic sign at integer values $n$ revealing an anti-correlation, i.e., an anti-persistence, of two ``successive'' increments $\delta x_1(t)$ and $\delta x_1(t+1)$.
This finding suggests an interpretation of the diffusion process as a time-discrete anti-persistent random walk, which will be specified in the next section.

\section{\label{sec:III}Anti-persistent random walk}

\begin{figure}
\includegraphics[width=\linewidth]{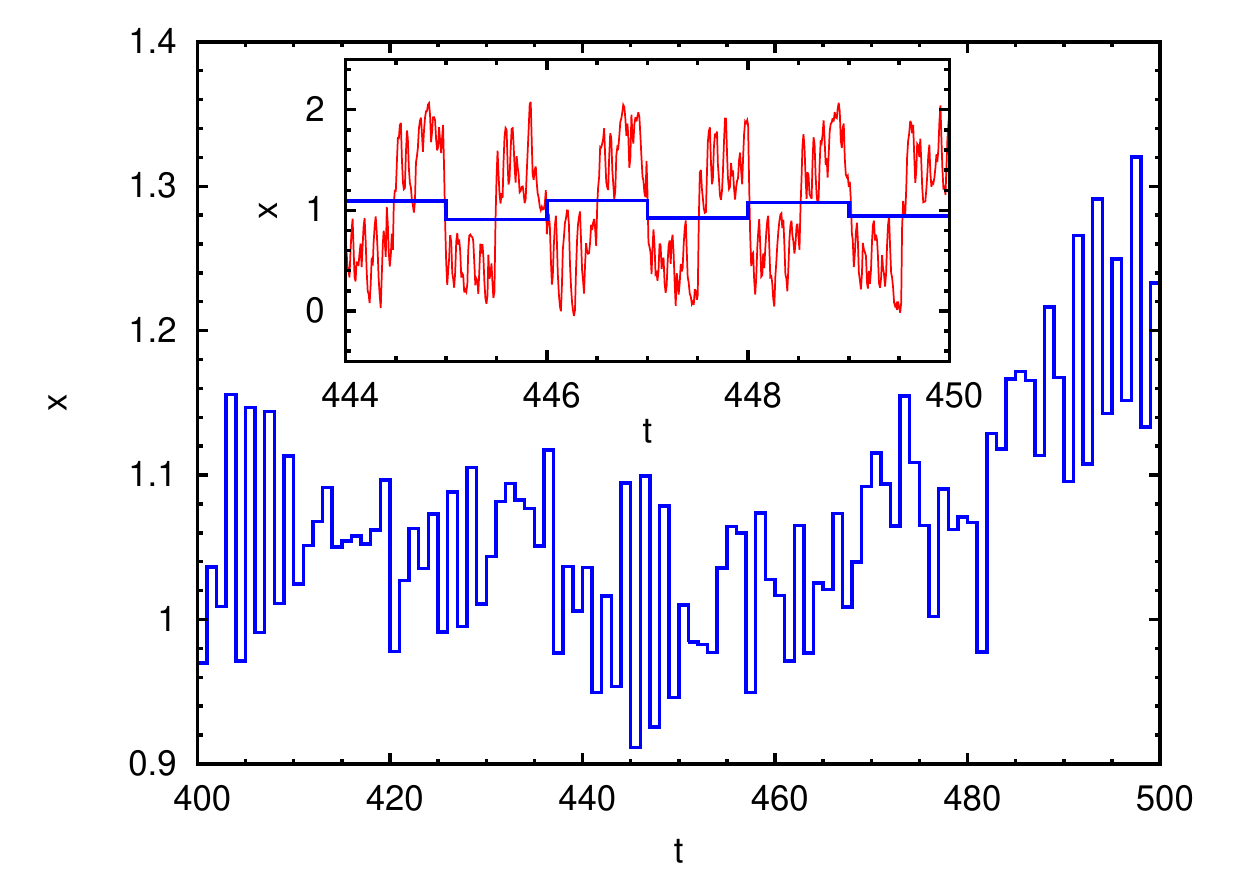}
\caption{\label{fig:center_of_mass}Time evolution of the center of mass per state interval (thick blue line) on a large time scale (main figure)
and on a short time scale (inset) compared with the corresponding solution $x(t)$ of the DDE (thin red line).
Same parameters as in Fig.~\ref{fig:trajectories}.}
\end{figure}

Motivated by the method of steps, which introduces a discretization in time of Eq.~(\ref{eq:DDE}) via the iteration of solution segments $x_n(t)$ defined on state intervals $[n-1,n]$,
and in order to get rid of the strong fluctuations per state interval, we consider another quantity that is able to capture the diffusive properties of our system very well,
namely the ``center of mass'' $S_n$ per state interval defined by
\begin{equation}
\label{eq:center_of_mass}
S_n=\int_{n-1}^nx(t)\,dt.
\end{equation}
By introducing increments $\delta S_n$ of this average via $\delta S_n=S_{n+1}-S_n$, the dynamics of the center of mass can be interpreted as a time-discrete random walk,
whose diffusion coefficient is determined by the statistics of its increments.
In the inset of Fig.~\ref{fig:center_of_mass}, we compare a typical solution of Eq.~(\ref{eq:DDE}) with the time evolution of its center of mass on a short time scale,
whereas the main figure shows the temporal behavior of the center of mass on a larger time scale.
The anti-persistence, i.e., a positive increment of the center of mass is more likely to be followed by a negative increment and vice versa, is clearly visible.

\begin{figure}
\includegraphics[width=\linewidth]{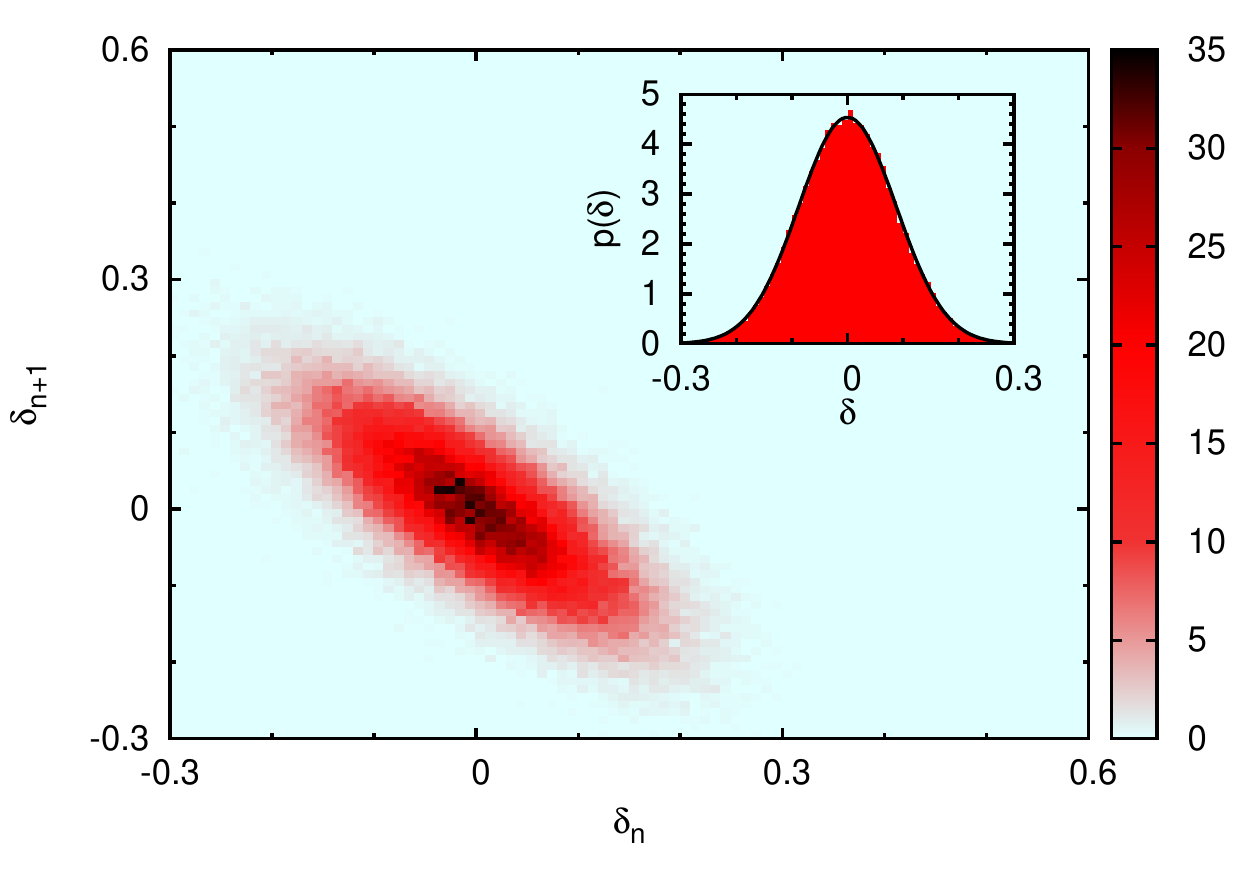}
\caption{\label{fig:increments_distribution}The two-dimensional probability density $p(\delta_n,\delta_{n+1})$ of two successive increments $\delta S_n$ and $\delta S_{n+1}$ of the center of mass per state interval visualizes the antipersistence (main figure),
and the one-dimensional probability density $p(\delta_n)$ is well described by a Gaussian distribution $\mathcal{N}(0,\sigma^2)$ (black line) with zero mean and variance $\sigma^2=\langle\delta S_n^2\rangle\approx0.0077$ (inset).
Same parameters as in Fig.~\ref{fig:trajectories}.}
\end{figure}

This behavior is confirmed in Fig.~\ref{fig:increments_distribution}, which shows the two-dimensional probability density $p(\delta_n,\delta_{n+1})$ of two successive increments $\delta S_n$ and $\delta S_{n+1}$.
We can see that, for instance, a large positive value of $\delta_n$ is typically connected with a large negative value of $\delta_{n+1}$ what leads to the observed anti-persistence.
The one-dimensional distribution $p(\delta_n)$ of the increments is Gaussian as shown in the inset of Fig.~\ref{fig:increments_distribution}.
As expected, the mean value $\langle\delta S_n\rangle$ of the increments is equal to zero leading to a pure diffusion process without any drift.

\begin{figure}
\includegraphics[width=\linewidth]{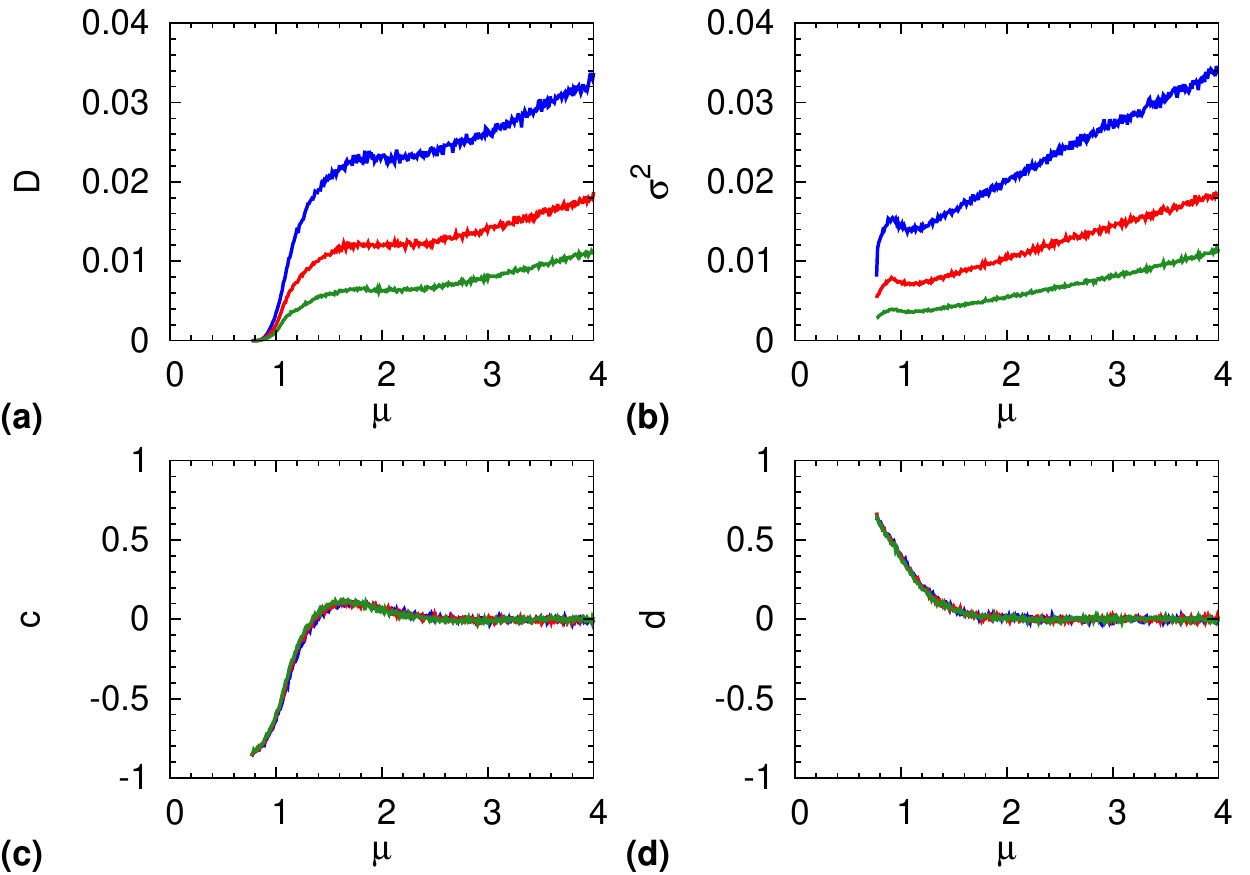}
\caption{\label{fig:increments_statistics}$\mu$-dependence of the diffusion coefficient $D$ of the DDE (a), the variance $\sigma^2$ of the increments $\delta S_n$ of the center of mass per state interval (b),
the correlation coefficient $c$ of two successive increments $\delta S_n$ and $\delta S_{n+1}$ (c), and the correlation coefficient $d$ of next-nearest increments $\delta S_n$ and $\delta S_{n+2}$ (d)
for three different values of $\Theta$ ($\Theta=25,50,100$ from top to bottom).
Panels (c) and (d) show that the correlation coefficients $c$ and $d$ do not depend on $\Theta$ asymptotically,
whereas the diffusion coefficient $D$ and the variance $\sigma^2$ in panels (a) and (b) are roughly proportional to $1/\Theta$.
($\mu>0.77$).}
\end{figure}

For a normal random walk, the diffusion coefficient is essentially determined by the variance $\sigma^2=\text{Var}(\delta S_n)=\langle\delta S_n^2\rangle=\text{Cov}(\delta S_n,\delta S_n)$ of the increments, $D=\sigma^2$.
For an anti-persistent random walk, however, correlations of the increments have to be taken into account.
For the following numerical and analytical considerations, we define the correlation coefficient $c$ of two successive increments by $c=\text{Cov}(\delta S_n,\delta S_{n+1})/\sigma^2$
and the correlation coefficient $d$ of next-nearest increments via $d=\text{Cov}(\delta S_n,\delta S_{n+2})/\sigma^2$.
We first start with a numerical investigation of these quantities in dependence on the nonlinearity parameter $\mu$ of the DDE, Eq.~(\ref{eq:DDE}) with Eq.~(\ref{eq:climbing-sine}), and the delay determined by the parameter $\Theta$.
In Figs.~\ref{fig:increments_statistics} (a) and (b), we compare the diffusion coefficient $D$ of the DDE and the variance $\sigma^2$ of the increments, respectively, in dependence on $\mu$ for three different values of $\Theta$.
A first observation is that both quantities roughly get halved if the value of $\Theta$ is doubled.
This is in agreement with a previous finding of the authors in \cite{albers2022}, where it was shown that the diffusion coefficient asymptotically vanishes as $D\sim1/\Theta$.
Furthermore, we can see that for larger values of $\mu$, the diffusion coefficient and the variance of the increments coincide,
whereas for smaller values of $\mu$, there are distinct discrepancies between these two quantities that can only be explained by taking the anti-persistence into account.
This is confirmed by looking at the correlation coefficients $c$ and $d$ in Figs.~\ref{fig:increments_statistics} (c) and (d), which are different from zero for smaller values of $\mu$, but go to zero for larger values of $\mu$.
In the former case, the correlation coefficient $c$ of two successive increments is negative, while the correlation coefficient $d$ of next-nearest increments is positive,
reflecting the anti-persistence of the increments in this parameter range.
Furthermore, we recognize that both correlation coefficients do not depend on $\Theta$.

In order to connect the dependence of the diffusion coefficient on the nonlinearity parameter with the $\mu$ dependence of the quantities $\sigma^2$, $c$, and $d$, we consider Markov models for the dynamics of the increments,
as it was successfully applied in modeling persistence effects in chaotic diffusion in extended two-dimensional billards \cite{gilbert_persistence_2009} and one-dimensional maps \cite{knight_capturing_2011}.
In the simplest case, a Markov process of zeroth order, where successive increments are completely independent from each other, the diffusion coefficient is just given by
\begin{equation}
\label{eq:D0}
D=\sigma^2
\end{equation}
as known from standard random walk theory.
The previous numerical results showed that this is only the case for larger values of $\mu$ where $c\approx d\approx0$.
As a next step, we consider a Markov process of first order for the dynamics of the increments.
The numerical results in Fig.~\ref{fig:increments_distribution} support that the probability density $p(\delta_n,\delta_{n+1})$ of two successive increments $\delta S_n$ and $\delta S_{n+1}$
is given by a two-dimensional Gaussian distribution,
\begin{equation}
\label{eq:distribution_2d}
p(\delta_n,\delta_{n+1})=\frac{1}{\sqrt{(2\pi)^2\text{det}(\boldsymbol{\Sigma})}}\exp\left(-\frac{1}{2}\boldsymbol{\delta}^T\boldsymbol{\Sigma}^{-1}\boldsymbol{\delta}\right)
\end{equation}
with $\boldsymbol{\delta}=(\delta_n,\delta_{n+1})^T$, and the covariance matrix reads
\begin{equation}
\label{eq:covariance_matrix_2d}
\boldsymbol{\Sigma}=\sigma^2\begin{pmatrix}1&c\\c&1\end{pmatrix}.
\end{equation}
By using the one-dimensional probability density $p(\delta_n)$ of the increments,
\begin{equation}
\label{eq:distribution_1d}
p(\delta_n)=\mathcal{N}_{\delta_n}(0,\sigma^2)=\frac{1}{\sqrt{2\pi\sigma^2}}\exp\left(-\frac{\delta_n^2}{2\sigma^2}\right),
\end{equation}
we can calculate the conditional probability density
\begin{equation}
\label{eq:propagator_2d}
p(\delta_{n+1}|\delta_n)=\frac{p(\delta_{n+1},\delta_n)}{p(\delta_n)}=\mathcal{N}_{\delta_{n+1}}\boldsymbol{(}c\delta_n,\sigma^2(1-c^2)\boldsymbol{)}
\end{equation}
of finding an increment $\delta S_{n+1}$ at discrete time $n+1$ given an increment $\delta S_n$ at time $n$.
This is the fundamental quantity that defines the Markov process of first order and is also known as propagator.
With the help of the propagator, one can determine all joint probability densities.
This allows us to calculate the covariance function of the increments,
\begin{equation}
\label{eq:covariance1}
\text{Cov}(\delta S_n,\delta S_0)=\langle\delta S_n\delta S_0\rangle=\int_{\mathbb{R}}\int_{\mathbb{R}}\delta_n\delta_0\,p(\delta_n,\delta_0)\,d\delta_n\,d\delta_0,
\end{equation}
where the joint probability density $p(\delta_n,\delta_0)$ is a marginal distribution of the overall probability density $p(\delta_n,\delta_{n-1},\dots,\delta_0)$ that can be expressed by the propagator leading to
\begin{equation}
\begin{split}
\label{eq:covariance2}
\langle\delta S_n\delta S_0\rangle=\int_{\mathbb{R}}\cdots\int_{\mathbb{R}}\delta_n\,p(\delta_n|\delta_{n-1})\,p(\delta_{n-1}|\delta_{n-2})\cdots\\
\cdots p(\delta_1|\delta_0)\,\delta_0\,p(\delta_0)\,d\delta_n\cdots d\delta_0.
\end{split}
\end{equation}
The $(n+1)$-fold integral on the right-hand side of Eq.~(\ref{eq:covariance2}) can be evaluated with the help of the propagator in Eq.~(\ref{eq:propagator_2d}),
\begin{equation}
\label{eq:covariance3}
\langle\delta S_n\delta S_0\rangle=\sigma^2c^n.
\end{equation}
From the covariance function of the increments, we can calculate the MSD,
\begin{equation}
\begin{split}
\label{eq:MSD1}
\langle(S_n-S_0)^2\rangle&=\left\langle\left(\sum\limits_{i=0}^{n-1}\delta S_i\right)^2\right\rangle=\sum\limits_{i=0}^{n-1}\sum\limits_{j=0}^{n-1}\langle\delta S_i\delta S_j\rangle\\
&=\sum\limits_{i=0}^{n-1}\langle\delta S_i^2\rangle+2\sum\limits_{i=1}^{n-1}\sum\limits_{j=0}^{i-1}\langle\delta S_i\delta S_j\rangle\\
&=\sigma^2n+2\sigma^2\sum\limits_{i=1}^{n-1}\sum\limits_{k=1}^ic^k.
\end{split}
\end{equation}
For the MSD of the center of mass, we obtain
\begin{equation}
\label{eq:MSD2}
\langle(S_n-S_0)^2\rangle=\frac{1+c}{1-c}\sigma^2n+2\sigma^2\frac{c^{n+1}-c}{(1-c)^2},
\end{equation}
and, therefore, for the diffusion coefficient of the DDE, we get
\begin{equation}
\label{eq:D1}
D=\frac{1+c}{1-c}\,\sigma^2,
\end{equation}
which coincides with the result for the anti-persistent random walk on a one-dimensional lattice considered in \cite{halpern_anti-persistent_1996}.
This formula contains the special case of the zeroth order Markov process for $c=0$.
Similarly, we can also consider a Markov process of second order, which takes the correlation coefficient $d$ of next-nearest increments into account.
The details of the definition of this process as well as the corresponding derivations are provided in Appendix~\ref{sec:A}.
Here, we only state the final analytical result, i.e., the diffusion coefficient in dependence on $\sigma^2$, $c$, and $d$,
\begin{equation}
\label{eq:D2}
D=\frac{1+c}{1-c}\,\frac{1+d-2c^2}{1-d}\,\sigma^2.
\end{equation}
This formula contains the special case of the first order Markov process for $d=c^2$.
Note that a similar expression for an anti-persistent random walk on a one-dimensional lattice was derived in \cite{gilbert2010}, where the diffusion coefficient depends on persistence probabilities.

\begin{figure*}
\includegraphics[width=\linewidth]{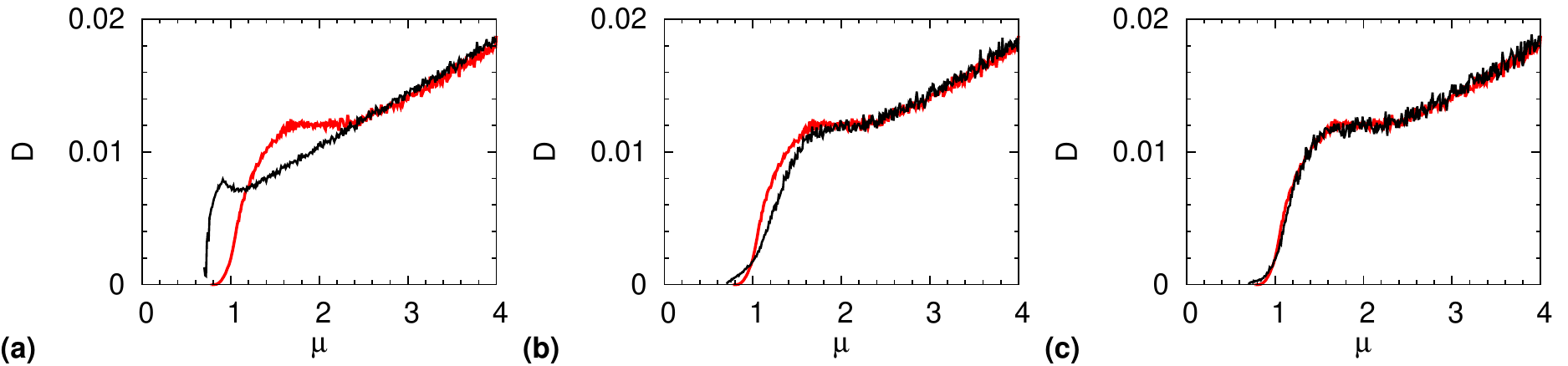}
\caption{\label{fig:diffusion_coefficient}Comparison of the numerically obtained diffusion coefficient $D$ of the DDE for different values of the nonlinearity parameter $\mu$ (red lines)
with the diffusion coefficients that are obtained for a Markov process of order zero (a), one (b), and two (c) (black lines) via Eqs.~(\ref{eq:D0},\ref{eq:D1},\ref{eq:D2})
from the numerically determined values for the variance $\sigma^2$ and the correlation coefficients $c$ and $d$ in Fig.~\ref{fig:increments_statistics}.
($\Theta=50$).}
\end{figure*}

In Fig.~\ref{fig:diffusion_coefficient}, we compare the numerically determined diffusion coefficient from the DDE
with the diffusion coefficient obtained by a Markov process of zeroth, first, and second order (from left to right) for the increments of the center of mass.
We thereby used Eq.~(\ref{eq:D0}), Eq.~(\ref{eq:D1}), and Eq.~(\ref{eq:D2}) with numerical values for $\sigma^2$, $c$, and $d$ from Fig.~\ref{fig:increments_statistics}.
As a final conclusion, we can state that whereas the Markov process of zeroth order is good enough to describe the diffusion coefficient for large values of the nonlinearity parameter $\mu$,
for smaller values of the parameter $\mu$, Markov processes of increasing order are needed.

\section{\label{sec:IV}Discussion and Summary}

\begin{figure}
\includegraphics[width=\linewidth]{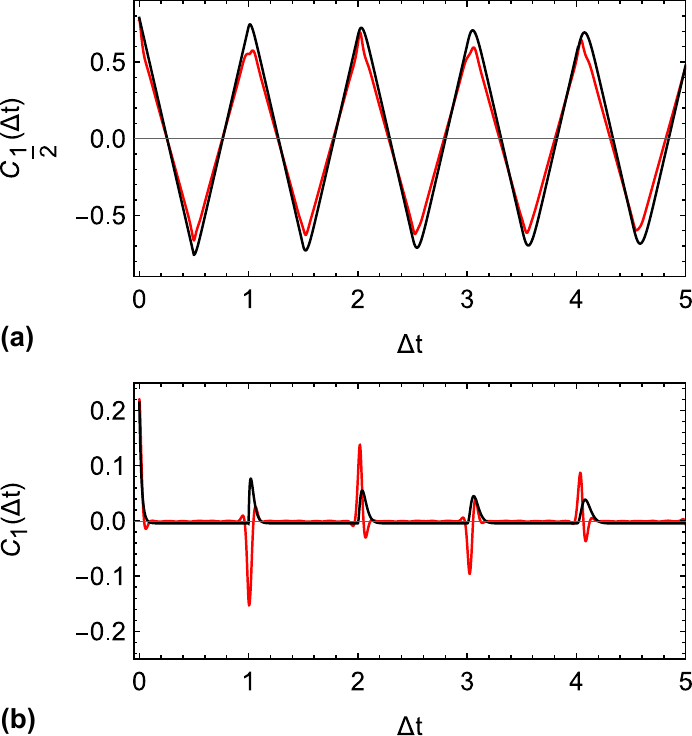}
\caption{\label{fig:covariance_function}The covariance function $C_{\eta}(\Delta t)=\langle\delta x_{\eta}(t)\delta x_{\eta}(t+\Delta t)\rangle$ of the increments $\delta x_{\eta}(t)=x(t+\eta)-x(t)$
of solutions $x(t)$ of the DDE, Eq.~(\ref{eq:DDE}), shows oscillations with slowly decreasing amplitude for $\eta=1/2$ (a) and sharp peaks of alternating algebraic sign for $\eta=1$ (b) (red curves).
The black lines are corresponding analytical results obtained from the stochastic delay differential equation, Eq.~(\ref{eq:SDDE}), with $\varsigma\approx0.035$ in (a) and $\varsigma\approx0.066$ in (b).
Same parameters as in Fig.~\ref{fig:trajectories} are used for Eq.~(\ref{eq:DDE}).}
\end{figure}

So far, we considered the anti-persistent random walk of the center of mass per state interval.
For a continuous-time dynamical system such as the DDE in Eq.~(\ref{eq:DDE}), however, there are several possible discretizations in time that can lead to different discrete-time random walks.
Let us consider again increments $\delta x_{\eta}(t)=x(t+\eta)-x(t)$ of solutions of Eq.~(\ref{eq:DDE}) and their covariance function defined by $C_{\eta}(\Delta t)=\langle\delta x_{\eta}(t)\delta x_{\eta}(t+\Delta t)\rangle$.
In Fig.~\ref{fig:covariance_function}, we compare the covariance functions $C_{1/2}(\Delta t)$ and $C_1(\Delta t)$.
We can see that whereas the covariance function $C_1(\Delta t)$ consists of a sequence of sharp peaks of alternating algebraic sign at integer values $n$,
the covariance function $C_{1/2}(\Delta t)$ is described by an oscillating function with a slowly decreasing amplitude.
$C_{1/2}(\Delta t)$ shows that there is a strong anti-persistence of these ``half increments'' for $\Delta t=1/2$.
The reason is the nearly periodic structure of chaotic solutions of Eq.~(\ref{eq:DDE}) with period equal to the constant delay as shown in the inset of Fig.~\ref{fig:center_of_mass}.
Note that there is a simple relation between both covariance functions, namely $C_1(\Delta t)=C_{1/2}(\Delta t-1/2)+2C_{1/2}(\Delta t)+C_{1/2}(\Delta t+1/2)$,
because the corresponding increments are related by $\delta x_1(t)=\delta x_{1/2}(t)+\delta x_{1/2}(t+1/2)$.
This means that the information contained in $C_1(\Delta t)$ is also included in $C_{1/2}(\Delta t)$ but not vice versa.
By using the approximation in Eq.~(\ref{eq:triangular}) for $C_{1/2}(\Delta t)$, one gets exactly zero for $C_1(\Delta t)$.
This demonstrates that $C_1(\Delta t)$ describes the deviations from the perfect triangular shape in Eq.~(\ref{eq:triangular}) that are hardly visible in Fig.~\ref{fig:covariance_function} (a).
Moreover, in Fig.~\ref{fig:covariance_function}, the anti-persistence in $C_1(\Delta t)$ is difficult to realize from $C_{1/2}(\Delta t)$.
This can also be seen in the inset of Fig.~\ref{fig:center_of_mass}.
The anti-persistence of the ``half increments'' is clearly visible due to the nearly periodicity of the solution $x(t)$, while the anti-persistence of the ``full increments'' only becomes visible by considering the center of mass.
The nearly periodic structures of the DDE seem to be a very robust phenomenon because they can also be observed in linear stochastic delay differential equations (SDDEs).
Replacing the term $\mu\sin\boldsymbol{(}2\pi x(t-1)\boldsymbol{)}$ in Eq.~(\ref{eq:DDE}) with Eq.~(\ref{eq:climbing-sine}) by Gaussian white noise $\xi(t)$ with $\langle\xi(t)\rangle=0$ and $\langle\xi(t)\xi(t')\rangle=\delta(t-t')$ leads to the SDDE
\begin{equation}
\label{eq:SDDE}
\frac{1}{\Theta}\dot{x}(t)=-x(t)+x(t-1)+\varsigma\xi(t).
\end{equation}
For this SDDE, one can derive the correlation functions $C_{1/2}(\Delta t)$ and $C_1(\Delta t)$ as numerically determined inverse Fourier transforms of the corresponding analytically derived power spectra, see Appendix~\ref{sec:B}.
These results are also displayed in Fig.~\ref{fig:covariance_function}.
We can see that while the SDDE reproduces the shape of the covariance function $C_{1/2}(\Delta t)$ very well, the anti-persistence of the covariance function $C_1(\Delta t)$ can not be reproduced by the SDDE.
The SDDE can explain the anti-persistence of the ``half increments'' because its solutions show the same nearly periodic structures,
but the anti-persistence of the ``full increments'' or the center of mass (also discussed in Appendix~\ref{sec:B}) is not captured by the SDDE.
In principle, one can derive the diffusion coefficient of the DDE from all covariances discussed so far, but we think, however, that the anti-persistent random walk defined in the previous section is the most natural one
and because of the suppression of the strong fluctuations per state interval due to the averaging over these state intervals, probably most suited to investigate the influence of anti-persistence on the diffusive properties of DDEs.

In summary, we have shown that chaotic diffusion appearing in a typical class of DDEs with a linear instantaneous and a nonlinear delayed term can be described by an anti-peristent random walk in a wide range of parameters.
We investigated the dependence of the anti-persistence on the strength of the nonlinearity and the delay and described the incremental process with Markov models.
With numerical and analytical considerations, we demonstrated that for large nonlinearities, the anti-persistence gets lost, and the increments are completely uncorrelated,
whereas for a decreasing strength of the nonlinearity, Markov processes of higher order are needed.
To the best knowledge of the authors, the occurrence of anti-persistent random walks in DDEs has never been reported before in the literature.

\appendix

\section{\label{sec:A}Derivation of Eq.~(\ref{eq:D2})}

In this appendix, we derive the diffusion coefficient of an unbiased anti-persistent random walk, $S_{n+1}=S_n+\delta S_n$, whose increments $\delta S_n$ follow a Markov process of second order with Gaussian probability densities.
Our objective is to obtain the diffusion coefficient $D$ in dependence on the variance $\sigma^2=\langle\delta S_n^2\rangle$ of the increments,
the correlation coefficient $c=\text{Cov}(\delta S_{n-1},\delta S_n)/\sigma^2$ of two successive increments, and the correlation coefficient $d=\text{Cov}(\delta S_{n-2},\delta S_n)/\sigma^2$ of next-nearest increments.
The derivation is analogous to the one in the main text for a Markov process of first order.
The distribution of three successive increments is given by the three-dimensional Gaussian probability density
\begin{equation}
\label{eq:distribution_3d}
p(\delta_n,\delta_{n-1},\delta_{n-2})=\frac{1}{\sqrt{(2\pi)^3\text{det}(\boldsymbol{\Sigma})}}\exp\left(-\frac{1}{2}\boldsymbol{\delta}^T\boldsymbol{\Sigma}^{-1}\boldsymbol{\delta}\right),
\end{equation}
where $\boldsymbol{\delta}=(\delta_n,\delta_{n-1},\delta_{n-2})^T$, and the covariance matrix is given by
\begin{equation}
\label{eq:covariance_matrix_3d}
\boldsymbol{\Sigma}=\sigma^2\begin{pmatrix}1&c&d\\c&1&c\\d&c&1\end{pmatrix}.
\end{equation}
The propagator, the conditional probability density of finding an increment $\delta S_n$ at discrete time $n$ given two increments $\delta S_{n-1}$ and $\delta S_{n-2}$ at times $n-1$ and $n-2$, respectively,
can be obtained from Eq.~(\ref{eq:distribution_3d}) with Eq.~(\ref{eq:covariance_matrix_3d}) and the two-dimensional distribution of the increments in Eq.~(\ref{eq:distribution_2d}),
\begin{equation}
\begin{split}
\label{eq:propagator_3d}
p(\delta_n|\delta_{n-1},\delta_{n-2})&=\frac{p(\delta_n,\delta_{n-1},\delta_{n-2})}{p(\delta_{n-1},\delta_{n-2})}\\[1ex]
&=\mathcal{N}_{\delta_n}(\alpha\delta_{n-1}+\beta\delta_{n-2},\gamma),
\end{split}
\end{equation}
where we used the abbreviations $\mathcal{N}_{\delta_n}(\mu,\sigma^2)=(2\pi\sigma^2)^{-1/2}\exp(-(\delta_n-\mu)^2/(2\sigma^2))$ of an one-dimensional Normal distribution and
\begin{equation}
\label{eq:abbreviations}
\alpha=\frac{c(1-d)}{1-c^2},\;\beta=\frac{d-c^2}{1-c^2},\;\gamma=\sigma^2\frac{(1+d-2c^2)(1-d)}{1-c^2}.
\end{equation}
Note that for a Markov process of first order, i.e., $d=c^2$, we recover the propagator in Eq.~(\ref{eq:propagator_2d}).
The covariance function of the increments is defined by
\begin{equation}
\label{eq:covariance4}
\text{Cov}(\delta S_n,\delta S_0)=\langle\delta S_n\delta S_0\rangle=\int_{\mathbb{R}}\int_{\mathbb{R}}\delta_n\delta_0\,p(\delta_n,\delta_0)\,d\delta_n\,d\delta_0,
\end{equation}
where the two-dimensional probability density $p(\delta_n,\delta_0)$ is the marginal distribution of the overall probability density $p(\delta_n,\dots,\delta_0,\delta_{-1})$.
The covariance function of $\delta S_n$ can be expressed by the propagator in Eq.~(\ref{eq:propagator_3d}) leading to
\begin{equation}
\begin{split}
\label{eq:covariance5}
\langle\delta S_n\delta S_0\rangle=\int_{\mathbb{R}}\cdots\int_{\mathbb{R}}\delta_n\,p(\delta_n|\delta_{n-1},\delta_{n-2})\,p(\delta_{n-1}|\delta_{n-2},\delta_{n-3})\\[1ex]
\cdots p(\delta_1|\delta_0,\delta_{-1})\,\delta_0\,p(\delta_0,\delta_{-1})\,d\delta_n\cdots d\delta_0\,d\delta_{-1}.
\end{split}
\end{equation}
We can calculate the $(n+2)$-fold integral by performing step by step the integrations with respect to $\delta_n$, $\delta_{n-1}$, and so on.
In the following, we consider the evolution of the prefactor in front of the product of propagators after performing $k$ integrations,
\begin{equation}
\begin{split}
\label{eq:integrand_implicit}
k=0&:\delta_{n}\\
k=1&:\alpha\delta_{n-1}+\beta\delta_{n-2}\\
k=2&:(\alpha^2+\beta)\delta_{n-2}+\alpha\beta\delta_{n-3}\\
k=3&:(\alpha^3+2\alpha\beta)\delta_{n-3}+(\alpha^2\beta+\beta^2)\delta_{n-4}
\end{split}
\end{equation}
In general, we can write after performing $k$ integrations,
\begin{equation}
\label{eq:integrand_explicit}
f_k\delta_{n-k}+\beta f_{k-1}\delta_{n-k-1},
\end{equation}
where the coefficients $f_k$ are the numbers of a generalized Fibonacci sequence given by
\begin{equation}
\label{eq:fibonacci_implicit}
f_k=\alpha f_{k-1}+\beta f_{k-2},\quad f_{-1}=0,\,f_0=1.
\end{equation}
This linear difference equation can be solved by the ansatz $f_k=\lambda^k$ leading to the explicit formula
\begin{equation}
\label{eq:fibonacci_explicit}
f_k=\frac{(\alpha+\sqrt{\alpha^2+4\beta})^{k+1}-(\alpha-\sqrt{\alpha^2+4\beta})^{k+1}}{2^{k+1}\sqrt{\alpha^2+4\beta}}.
\end{equation}
From Eq.~(\ref{eq:integrand_explicit}) for $k=n$, we obtain for the covariance function of the increments
\begin{equation}
\begin{split}
\label{eq:covariance6}
\langle\delta S_n\delta S_0\rangle&=\int_{\mathbb{R}}\int_{\mathbb{R}}(f_n\delta_0+\beta f_{n-1}\delta_{-1})\delta_0\,p(\delta_0,\delta_{-1})\,d\delta_0\,d\delta_{-1}\\[1ex]
&=\sigma^2f_n+\sigma^2c\beta f_{n-1}.
\end{split}
\end{equation}
By using Eq.~(\ref{eq:MSD1}), we get for the MSD of the anti-persistent random walk
\begin{equation}
\label{eq:MSD3}
\langle(S_n-S_0)^2\rangle=\sigma^2n+2\sigma^2\sum\limits_{i=1}^{n-1}\sum\limits_{k=1}^if_k+c\beta f_{k-1}.
\end{equation}
The evaluation of the double sum on the right-hand side of Eq.~(\ref{eq:MSD3}) with the explicit formula in Eq.~(\ref{eq:fibonacci_explicit}) and the abbreviations in Eq.~(\ref{eq:abbreviations})
leads to the diffusion coefficient of the anti-persistent random walk, i.e., the asymptotic linear slope of its MSD,
\begin{equation}
\label{eq:D}
D=\frac{1+c}{1-c}\,\frac{1+d-2c^2}{1-d}\,\sigma^2.
\end{equation}
Because a single dicrete time step of the anti-persistent random walk of the center of mass per state interval corresponds to the iteration of one solution segment of length unity of the DDE,
we obtain Eq.~(\ref{eq:D2}) for the diffusion coefficient of the DDE.

\section{\label{sec:B}Covariance functions of the increments}

In this appendix, the covariance functions $C_{\eta}(\Delta t)$ of the increments $\delta x_{\eta}(t)=x(t+\eta)-x(t)$ with $\eta\in\{1/2,1\}$
and the increments $\delta S_n=S_{n+1}-S_n$ of the center of mass are analyzed in the limit $\Theta\rightarrow\infty$ for the SDDE~(\ref{eq:SDDE}).
Beyond the numerical estimation from ensembles of time series using the definition of the covariance function $C_{\eta}(\Delta t)=\langle\!\langle\delta x_{\eta}(t)\,\delta x_{\eta}(t+\Delta t)\rangle\!\rangle$,
for this system, there are at least four possible approaches to compute or estimate $C_{\eta}(\Delta t)$.
The first three approaches directly follow from the definition of the covariance function.
Assuming stationarity of the increments $\delta x_{\eta}(t)$, one has
\begin{equation}
\begin{split}
C_{\eta}(\Delta t)&=C(t+\eta,t+\Delta t+\eta)+C(t,t+\Delta t)\\
&\quad-C(t+\eta,t+\Delta t)-C(t,t+\Delta t+\eta),
\end{split}
\end{equation}
where $C(t,t')$ is the covariance function of $x(t)$, which can be obtained via the eigen mode expansion of the deterministic part of the SDDE~(\ref{eq:SDDE}) as shown in \cite{amann_basic_2007},
or using the analytical expression for the Green function as shown in \cite{budini2004}.
A third approach may be derived from the method in \cite{porte_autocorrelation_2014}, where it is shown that the correlation function of $x(t)$ is a special solution of a certain deterministic DDE while requiring stationarity of the system.
However, in our case, stationarity can only be assumed for the increments $\delta x_{\eta}(t)$ but not for $x(t)$ since the considered system shows diffusion.
The fourth approach, which is the one we use in the following analysis, uses the fact that the covariance function of a random variable is given by the inverse Fourier transform of the power spectrum of the random variable,
which is known as Wiener–Khinchin theorem \cite{wiener_generalized_1930,khintchine_korrelationstheorie_1934}.
Since the power spectrum $S_{\eta}(\omega)$ of $\delta x_{\eta}(t)$ is connected to the power spectrum $S(\omega)$ of $x(t)$ by $S_{\eta}(\omega)=2(1-\cos(\eta\omega))S(\omega)$, $S_{\eta}(\omega)$ is given by
\begin{equation}
\label{eq:pws}
S_{\eta}(\omega)=\frac{1-\cos(\eta\omega)}{1-\cos(\omega)+\frac{\omega}{\Theta}\left(\frac{1}{2}\frac{\omega}{\Theta}+\sin(\omega)\right)}\,\varsigma^2,
\end{equation}
where $S(\omega)$ was obtained from the Fourier transform of the SDDE~(\ref{eq:SDDE}) according to \cite{budini2004}.
The covariance functions $C_{\eta}(\Delta t)$ for $\eta=1/2$ and $\eta=1$ shown in Fig.~\ref{fig:covariance_function} were computed numerically by approximating the inverse Fourier transform of Eq.~\eqref{eq:pws} via a fast Fourier transform,
where for each $\eta$ a $\varsigma$ was chosen such that the resulting covariance functions for the SDDE~(\ref{eq:SDDE}) coincide with the numerical estimates of the covariance functions for the DDE~\eqref{eq:DDE} at $\Delta t=0$.

In the limit of large $\Theta$, $S_{1/2}(\omega)$ is large in the vicinity of $\omega\approx\omega_k=2\pi k[1-(\Theta+1)^{-1}]$ with $|k|=1,3,5,\dots$ and is negligible elsewhere.
$S_{1/2}(\omega)$ can be approximated by a sum of these peaks, which leads to
\begin{equation}
S_{1/2}(\omega)\approx\sum_{k\neq 0\text{ odd}}\frac{\frac{2\varsigma^2(\Theta+1)^2}{\pi k^2}}{\pi\frac{1}{2}\left(\frac{2\pi k}{(\Theta+1)}\right)^2\left[1+\left(\frac{\omega-\omega_k}{\frac{1}{2}\left(\frac{2\pi k}{(\Theta+1)}\right)^2}\right)^2\right]}
\end{equation}
and is in agreement with the observation made in \cite{amann_basic_2007} that the power spectrum essentially is a sum of Lorentzians.
The summands were derived by approximating the denominator of the fraction on the right-hand side of Eq.~(\ref{eq:pws}) by its Taylor series at $\omega=2\pi k$, while dropping terms with an order larger than $\omega^4$.
The resulting polynomial is minimized by $\omega\approx\omega_k=2\pi k[1-(\Theta+1)^{-1}]$ and can be approximated in the vicinity of $\omega_k$ by a second order polynomial in the limit of large $\Theta$.
The numerator was approximated by $1-\cos(\omega/2)\approx2$.
$C_{1/2}(\Delta t)$ is obtained by applying the inverse Fourier transform, which gives
\begin{equation}
\begin{split}
C_{1/2}(\Delta t)\approx\sum_{k=1,3,5,\dots}\frac{2\varsigma^2(\Theta+1)^2}{(\pi k)^2}e^{-\frac{1}{2}\left(\frac{2\pi k}{(\Theta+1)}\right)^2|\Delta t|}\\[1ex]
\times\cos(2\pi k(1-(\Theta+1)^{-1})\Delta t).
\end{split}
\end{equation}
For $|\Delta t|\rightarrow0$ or $\Theta\rightarrow\infty$, one has
\begin{equation}
\begin{split}
\label{eq:triangular}
C_{1/2}(\Delta t)\approx\frac{\varsigma^2(\Theta+1)^2}{4}\hspace{11em}\\[1ex]
\times\left(1-\frac{2}{\pi}\arccos(\cos(2\pi(1-(\Theta+1)^{-1})\Delta t))\right),
\end{split}
\end{equation}
which confirms the triangular shape observed in Fig.~\ref{fig:covariance_function}.
For $|\Delta t|\gg\Theta^2$, one has
\begin{equation}
\begin{split}
C_{1/2}(\Delta t)\approx\frac{2\varsigma^2(\Theta+1)^2}{\pi^2}e^{-\frac{1}{2}\left(\frac{2\pi}{(\Theta+1)}\right)^2|\Delta t|}\hspace{4em}\\[1ex]
\times\cos(2\pi(1-(\Theta+1)^{-1})\Delta t).
\end{split}
\end{equation}

For $\eta=1$, the previous approximation approach is not suitable, since, in this case, the background of $S_{\eta}(\omega)$ is not negligible compared to the peaks at $\omega\approx\omega_k=2\pi k[1-(\Theta+1)^{-1}]$ with $|k|=1,2,3,\dots$.
Nevertheless, the covariance function $\bar{C}_1(\Delta n)=\langle\delta S_n\delta S_{n+\Delta n}\rangle=\langle\delta S_0\delta S_{\Delta n}\rangle$ of the increments $\delta S_n$ of the center of mass can be derived from Eq.~(\ref{eq:pws}) in the limit $\Theta\rightarrow\infty$.
Therefore, it can be shown via a straight forward calculation that $\bar{C}_1(\Delta n)$ is connected to the covariance function $C_1$ of the increments $\delta x_{\eta}(t)$ by
\begin{equation}
\bar{C}_1(\Delta n)=\int_{-1}^1du\,(1-|u|)\,C_1(\Delta n+u),
\end{equation}
where only the definitions of the increments, the center of mass, Eq.~(\ref{eq:center_of_mass}), and the covariance function are used.
Inserting the inverse Fourier transform of Eq.~(\ref{eq:pws}) for $C_1$ and performing the integral over $u$ gives
\begin{equation}
\bar{C}_1(\Delta n)=\frac{2\varsigma^2}{\pi}\int_0^{\infty}d\omega\,\frac{(1-\cos(\omega))^2\,\cos(\Delta n\,\omega)}{\omega^2\left(1-\cos(\omega)+\frac{\omega}{\Theta}\left(\frac{1}{2}\frac{\omega}{\Theta}+\sin(\omega)\right)\right)}.
\end{equation}
In the limit $\Theta\rightarrow\infty$, we obtain
\begin{equation}
\begin{split}
\bar{C}_1(\Delta n)&=\frac{2\varsigma^2}{\pi}\int_0^{\infty}d\omega\,\frac{(1-\cos(\omega))\,\cos(\Delta n\,\omega)}{\omega^2}\\
&=\begin{cases}\varsigma^2&\text{if }\Delta n=0\\0&\text{else}\end{cases}.
\end{split}
\end{equation}
As a result, the correlation of successive increments $\delta S_n$ vanishes in the limit $\Theta\rightarrow\infty$ and thus, for the system governed by the SDDE~(\ref{eq:SDDE}),
the random walk given by the center of mass $S_n$ is not anti-persistent in this limit.

\vspace{1ex}

\begin{acknowledgments}
The authors gratefully acknowledge funding by the Deutsche Forschungsgemeinschaft (DFG, German Research Foundation) - 438881351;  456546951.
\end{acknowledgments}

\bibliography{references}

\end{document}